# Causal Discovery of Linear Cyclic Models from Multiple Experimental Data Sets with Overlapping Variables

**Antti Hyttinen**
HIIT & Dept. of Computer Science
University of Helsinki
Finland

**Frederick Eberhardt**
Dept. of Philosophy
Carnegie Mellon University
Pittsburgh, PA, USA

**Patrik O. Hoyer**
HIIT & Dept. of Computer Science
University of Helsinki
Finland

## Abstract

Much of scientific data is collected as randomized experiments intervening on some and observing other variables of interest. Quite often, a given phenomenon is investigated in several studies, and different sets of variables are involved in each study. In this article we consider the problem of integrating such knowledge, inferring as much as possible concerning the underlying causal structure with respect to the union of observed variables from such experimental or passive observational overlapping data sets. We do not assume acyclicity or joint causal sufficiency of the underlying data generating model, but we do restrict the causal relationships to be linear and use only second order statistics of the data. We derive conditions for full model identifiability in the most generic case, and provide novel techniques for incorporating an assumption of faithfulness to aid in inference. In each case we seek to establish what is and what is not determined by the data at hand.

## 1 INTRODUCTION

For many scientific domains different research groups study the same or closely related phenomena. In general the resulting data sets may contain different sets of variables with only partial overlap, and some of the data sets may be passive observational, while others have resulted from experiments intervening on a subset of the variables. In the context of discovering causal pathways, it would be desirable to be able to integrate all the findings of these 'overlapping' experimental and observational data sets, not only to gain an overall view, but, where possible, to also detect causal relations that were not identified in any individual study.

In the literature on causal discovery methods this problem has been addressed in two complementary, yet so far, independent ways. On the one hand a variety of algorithms have been developed that integrate data from *overlapping* but purely *passive observational* data sets (Tillman et al., 2009; Triantafillou et al., 2010; Tillman and Spirtes, 2011). On the other hand there are several procedures that combine multiple *experimental* studies on the *same* set of variables (Cooper and Yoo, 1999; Tong and Koller, 2001; Murphy, 2001; Eaton and Murphy, 2007; He and Geng, 2008; Schmidt and Murphy, 2009; Eberhardt et al., 2010; Hyttinen et al., 2010).

In this article we present methods that combine the two approaches: we consider data sets that contain passive observational or experimental measures of overlapping sets of variables. We do *not* assume that the set of variables is jointly causally sufficient; in other words we allow for the possible existence of common causes of the measured variables that are not measured in *any* of the available data sets. Moreover, unlike any of the previous work on overlapping data sets, we do *not* assume that the underlying causal structure is acyclic. To ensure that we have a well-defined representation of the *cyclic* relations, we assume that the causal relations are *linear* (see Section 2). The assumption of linearity also enables the use of quantitative constraints that can lead to the identifiability of the underlying causal structure in cases which are non-identifiable without such a parametric assumption (Hyttinen et al., 2011). We provide necessary and sufficient conditions under which the true causal model can be identified, but given the weak set of background assumptions, these are inevitably very demanding. In general there may only be a very limited number of experimental data sets, and so the remaining underdetermination can still be quite substantial. We then provide a novel approach for incorporating the assumption of *faithfulness* (Spirtes et al., 1993) to aid in model search for this family of linear models. We show that, for sparse graphs, a significant amount of structure can be inferred using this framework.

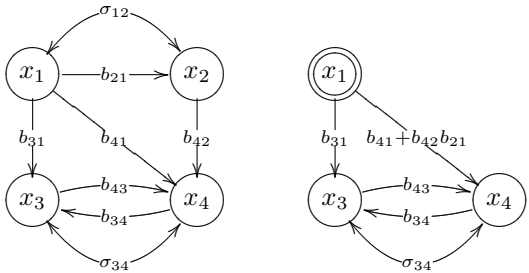

Figure 1: An example of a linear cyclic model with latent variables (left) and the corresponding manipulated model in an experiment where $x_1$ is intervened, $x_3$, $x_4$ are observed and $x_2$ is hidden (right). Single-headed arrows denote non-zero direct effects while double-headed arrows represent the existence of latent confounders. The disturbance variables are omitted to improve readability.

This paper is structured as follows. We formalize the problem in Section 2, by defining the model family we are considering, specifying the representation used for overlapping data sets, and formulating the inference problem. Then, in Section 3, we derive necessary and sufficient conditions for the identifiability of the model, in the absence of extra assumptions. In Section 4, we consider an approach to integrate simple faithfulness constraints, which we then extend to a more general and powerful method in Sections 5-6. Simulations are provided in Section 7, and some conclusions in Section 8.

## 2 MODEL AND PROBLEM DEFINITION

We use the following formal representation of the setting with overlapping data sets: Given data sets $\mathcal{D}_1, \ldots, \mathcal{D}_K$ with measurements on the sets of variables $\mathcal{V}_1, \ldots, \mathcal{V}_K$, respectively, let the joint set of variables be $\mathcal{V} = \{x_1, \ldots, x_n\} = \bigcup_{k=1,\ldots,K} \mathcal{V}_k$. Obviously, the individual sets $\mathcal{V}_k$ of variables in $\mathcal{D}_k$ will generally not be causally sufficient, but we also allow the *joint* set of variables $\mathcal{V}$ to be causally insufficient, i.e. there can be confounders that are not measured in any of the available data sets. Each data set $\mathcal{D}_k$ is considered to be the result of an experiment $\mathcal{E}_k$ that partitions the variables $\mathcal{V}$ into $\mathcal{J}_k$ (intervened), $\mathcal{U}_k$ (observed), and $\mathcal{L}_k$ (hidden) variables. We assume that all variables in $\mathcal{V}$ are measured in at least one of the experiments, i.e. for any $x \in \mathcal{V}$ there exists some experiment $\mathcal{E}_k$ with $1 \leq k \leq K$ such that $x \in \mathcal{J}_k \cup \mathcal{U}_k$. Passive observational studies are trivially represented by 'null'-experiments where $\mathcal{J}_k = \emptyset$.

We assume that there is a single generating model over the variables in $\mathcal{V}$ that, suitably manipulated for each experiment, gives rise to all the data sets. We consider the model class of linear cyclic models with latent variables (linear non-recursive SEMs with correlated disturbances) (Bollen, 1989). The model can be described by the equation

$$\mathbf{x} := \mathbf{Bx} + \mathbf{e} \qquad (1)$$

where $\mathbf{x}$ is a vector that denotes the variables in $\mathcal{V}$ while $\mathbf{e}$ represents the corresponding exogenous disturbances. The model is parameterized by a direct effects matrix $\mathbf{B}$ and a covariance matrix of the disturbances $\mathrm{cov}(\mathbf{e}) = \mathbf{\Sigma_e}$. Confounding due to variables *external* to the joint set of variables $\mathcal{V}$ is modeled implicitly by non-zero off-diagonal entries in $\mathbf{\Sigma_e}$ that correlate the disturbances. Figure 1 shows how the structure of such a model can be described by a mixed graph. Note that we assume throughout that the diagonal of $\mathbf{B}$ is zero; this is not a substantial assumption.[1]

We use the standard notion of "surgical" interventions (Pearl, 2000), where an intervention makes the intervened variable independent of its normal causes. Formally, the intervened model $(\tilde{\mathbf{B}}, \tilde{\mathbf{\Sigma}}_\mathbf{e})$ is the same as the unintervened model $(\mathbf{B}, \mathbf{\Sigma_e})$ except that each row in $\tilde{\mathbf{B}}$ corresponding to edges incident on some $x_i \in \mathcal{J}$ is set to zero (thereby breaking the edges into any such variables); similarly, for the $i$th row and column of $\tilde{\mathbf{\Sigma}}_\mathbf{e}$. An intervention vector $\mathbf{c}$ is added to the model Equation (1) that determines the values of (only) the intervened variables according to an intervention distribution with $\boldsymbol{\mu_c}$ and $\mathbf{\Sigma_c}$. For notational simplicity we assume that the intervened variables are randomized independently with zero mean and unit variance. This does not seriously limit the applicability of the methods.[2] For any individual (experimental) data set $\mathcal{D}_k$ we then assume that only the values of $\mathbf{x}$ corresponding to variables in $\mathcal{J}_k \cup \mathcal{U}_k$ are revealed (see Figure 1, right).

To ensure that the cyclic model has a well-defined equilibrium distribution, Hyttinen et al. (2012) examined models that are *weakly stable*, that is, models for which the $\mathbf{I} - \mathbf{B}$ matrix of the model (where $\mathbf{I}$ is the identity matrix) and all corresponding manipulated versions are invertible. Weak stability implies that the observed distribution follows a covariance matrix $\mathrm{cov}(\mathbf{x}) = (\mathbf{I} - \mathbf{B})^{-1} \mathbf{\Sigma_e} (\mathbf{I} - \mathbf{B})^{-T}$. However, in the

---
[1] There is inherent underdetermination of the diagonal elements of $\mathbf{B}$, related to self-loops (edges from a node to itself) in the model. These do not affect the equilibrium distribution in linear cyclic models, and can be handled by a suitable normalization (Hyttinen et al., 2012).

[2] A detailed formal account of the intervened model is given by Hyttinen et al. (2012), see their Equation 7 and Lemma 5.

case of overlapping data sets we have to make a slightly stronger assumption to obtain the desired identifiability results:

**Assumption 1 (No unit cycles)** *The sum-product of edge-coefficients on any subset of paths from a variable back to itself cannot sum up to exactly 1.*

Weak stability follows from this assumption. The proof of this claim, and all subsequent results in this paper, can be found in the Appendix in the Supplementary Material.

We can now state our learning problem formally: Given data sets $\mathcal{D}_1, \ldots, \mathcal{D}_K$ containing measurements of the marginalized distribution of a (experimentally manipulated) linear cyclic model with latent variables $(\mathbf{B}, \mathbf{\Sigma_e})$ that satisfies Assumption 1, identify as many causal relations among the joint set of variables, i.e. entries in $\mathbf{B}$, as possible.

## 3 IDENTIFIABILITY

One of the advantages of linear cyclic models is that the correlation measured between an intervened variable $x_i$ and a non-intervened variable $x_j$ corresponds to the sum-product of all the directed paths from $x_i$ to $x_j$ in a graph where all arcs into the intervened variables in $\mathcal{J}_k$ are cut. Eberhardt et al. (2010) refer to these terms as *experimental effects* and use the notation $\text{cov}_k(x_i, x_j) = t(x_i \leadsto x_j || \mathcal{J}_k)$. When the intervention set of an experimental effect contains all but one variable, then it corresponds to the *direct effect*: $t(x_i \leadsto x_j || \mathcal{V} \setminus \{x_j\}) = b(x_i \to x_j) = \mathbf{B}[j, i]$. Similarly, when only one variable is subject to intervention, we have the *total effect*: $t(x_i \leadsto x_j || \{x_i\}) = t(x_i \leadsto x_j)$.

Eberhardt et al. (2010) describe a learning algorithm which uses experimental effects estimated from the data to form linear equations on the unknown total effects. In particular, for an experiment with a variable $x_i \in \mathcal{J}_k$ and a variable $x_u \in \mathcal{U}_k$, one can obtain the constraint

$$t(x_i \leadsto x_u) = \sum_{x_j \in \mathcal{J}_k \setminus \{x_i\}} t(x_i \leadsto x_j) t(x_j \leadsto x_u || \mathcal{J}_k) \quad (2)$$
$$+ t(x_i \leadsto x_u || \mathcal{J}_k).$$

Note that for all $x_j \in \mathcal{J}_k$ the experimental effects $t(x_j \leadsto x_u || \mathcal{J}_k)$ are directly given by the relevant covariances (see above) in the data from this experiment, so the equation is linear in the unknown total effects. The intuition behind the equation is that the set of all directed paths from $x_i$ to $x_u$ in the original (unmanipulated) model can be separated into those going through each of the other intervened variables (the ones in $\mathcal{J}_k \setminus \{x_i\}$) and the remaining ones. Collecting a sufficient number of such linear constraints makes it possible to identify all total effects, from which it is straightforward to obtain the direct effects, as described in their paper.

While their learning procedure was constructed for the fully observed case, we note that it is directly applicable to our overlapping data sets case, since for any $x_i \in \mathcal{J}_k$ and any $x_u \in \mathcal{U}_k$, all required experimental effects are covariances between observed variables (and do not involve any variables in $\mathcal{L}_k$). Moreover, the identifiability results described in Eberhardt et al. (2010) and Hyttinen et al. (2012) can be generalized to the current setting once it is clarified how their conditions apply when the set of variables $\mathcal{V}$ is split into three sets $\mathcal{J}_k$, $\mathcal{U}_k$ and $\mathcal{L}_k$, rather than just $\mathcal{J}_k$ and $\mathcal{U}_k$. For this purpose, we say that a given set of experiments $\mathcal{E}_1, \ldots, \mathcal{E}_K$ satisfies the *pair condition* for the ordered pair of variables $(x_j, x_u)$ if there is an experiment $\mathcal{E}_k$, with $1 \leq k \leq K$, such that $x_j \in \mathcal{J}_k$ and $x_u \in \mathcal{U}_k$. Note that an experiment with $x_j \in \mathcal{J}_k$ and $x_l \in \mathcal{L}_k$ does not satisfy the pair condition for the pair $(x_j, x_l)$ even though $x_l$ is not subject to intervention. The identifiability conditions for overlapping data sets are then as follows:

**Theorem 1 (Sufficiency)** *Given some set of experiments, a linear cyclic model with latent variables satisfying Assumption 1 is fully identified if the pair condition is satisfied for all ordered pairs $(x_i, x_j) \in \mathcal{V} \times \mathcal{V}$, $x_i \neq x_j$.*

Note that unlike in Eberhardt et al. (2010), our sufficiency result here requires the slightly stronger Assumption 1. We further show that the sufficient condition is also in the worst case necessary.

**Theorem 2 (Worst Case Necessity)** *Given any set of experiments that does* not *satisfy the pair condition for all ordered pairs of variables $(x_i, x_j) \in \mathcal{V} \times \mathcal{V}$, $x_i \neq x_j$, there exist two distinct linear cyclic models with latent variables satisfying Assumption 1 that are indistinguishable given those experiments.*

So with minor adjustments to the conditions and assumptions, we obtain for the overlapping data sets case very similar identifiability results as were given for the case when all data sets shared the same set of variables. In fact, as indicated in Appendix C, the learning algorithm of Eberhardt et al. (2010) is also *complete* given the background assumptions we have been considering here, in the sense that the information gained from the experimental effects fully exhausts what can be inferred about the underlying causal model. That is, correlations between non-intervened variables in this case provide no extra benefit. In Section 7 we use this algorithm ('EHS') as a baseline comparison of what a complete procedure can infer without making stronger search space assumptions. Together, Theorems 1 and 2 provide *a priori* guidelines to identify which experimental results are needed to complement any available ones. Note that $K$ specifically chosen experiments are enough to the satisfy the identifiability condition for models with up to $\binom{K}{\lfloor K/2 \rfloor}$ variables (Spencer, 1970).

## 4 FAITHFULNESS & BILINEARITY

In most realistic contexts the set of available overlapping data sets will not contain a sufficient variety of experimental interventions to satisfy the demanding identifiability condition given in the previous section. Hence, the full model will typically not be fully identified by the 'EHS' procedure. Furthermore, in most cases the overwhelming majority of the direct effects will remain underdetermined, as is evident from the simulations in Section 7.

Thus, a useful approach might be to strengthen the assumptions regarding the underlying model. The natural assumption, prevalent throughout causal discovery research, is the assumption of *faithfulness* (Spirtes et al., 1993), which can be seen as a simplicity or minimality assumption for causal discovery procedures. Most commonly used for inference in passive observational data, we consider the natural extension of the faithfulness assumption to experimental contexts: A data-generating model is said to be *faithful* if *all independences true in any (manipulated) observed distribution are consequences of the structure of the (manipulated) graph, and not due to the specific parameter values of the model.* In short, faithfulness enables us to infer from a statistical independence that there is an absence of a causal connection between the variables in question, in the particular context under which the statistical independence occurs. It is well known that heavy confounding often results in causal models that are effectively unfaithful given the limited sample size of the data. Nevertheless, if one has reason to believe that models are sparse and confounding limited, then faithfulness may constitute a reasonable assumption that can provide further insights.

Hyttinen et al. (2010) considered a set of simple faithfulness rules by which marginal or conditional independencies in the observed variables were used to derive constraints of the form $b(x_i \to x_j) = \mathbf{B}[j,i] = 0$, i.e. based on statistical independencies in the experimental data, some direct effects were inferred to equal zero. While these constraints are (trivially) linear in the direct effects, they are highly *non-linear* in the total effects, and thus cannot easily be integrated with linear constraints on the total effects of the form of Equation 2. To solve this problem Hyttinen et al. (2010) noted that one could, in fact, use the measured experimental effects to put linear constraints on the *direct effects*, as opposed to the total effects.

$$t(x_i \rightsquigarrow x_u || \mathcal{J}_k) = b(x_i \rightsquigarrow x_u) + \sum_{x_j \in \mathcal{V} \setminus (\mathcal{J} \cup \{x_u\})} t(x_i \rightsquigarrow x_j || \mathcal{J}_k) b(x_j \rightsquigarrow x_u) \quad (3)$$

In this way, the constraints from the experimental effects could be straightforwardly combined with faithfulness constraints, resulting in a fully linear system of equations constraining the direct effects (Hyttinen et al., 2010). The resulting linear system can be analyzed to determine which direct effects are identified and which are underdetermined; this procedure is empirically evaluated in Section 7 under the abbreviation 'HEH'.

Unfortunately, in the overlapping variables scenario, in most cases one cannot obtain many linear constraints on the direct effects. This is because one or more of the non-intervened variables may be latent, and one needs the experimental effects to *all* non-intervened variables in a given experiment to construct the linear constraints on the direct effects. In general, the experimental effects $t(x_i \rightsquigarrow x_j || \mathcal{J}_k)$ are not available when $x_j \in \mathcal{L}_k$, unless they can be inferred from the combination of results from all data sets. On the other hand, as we discussed in Section 3, we *can* obtain linear constraints on the *total* effects, because latents do not form a problem in this case. The remaining problem is thus that, in general, we may have a large set of linear constraints on the elements of the total effects matrix $\mathbf{T} = (\mathbf{I} - \mathbf{B})^{-1}$, where $\mathbf{T}[j,i] = t(x_i \rightsquigarrow x_j)$, combined with the knowledge that certain elements of $\mathbf{B}$ are zero, but finding a solution to this combined set of constraints is no longer trivial.

One approach to solving this problem is to take the *combination* of the direct effects (elements of $\mathbf{B}$) *and* the total effects (elements of $\mathbf{T}$) as the unknown free parameters of the problem. In this way, we now have a *bilinear* system of equations: There are equations linear in the elements of $\mathbf{B}$, equations linear in the elements of $\mathbf{T}$, and the constraint $\mathbf{T}(\mathbf{I} - \mathbf{B}) = \mathbf{I}$ yields equations that are bilinear; they are linear in one parameter vector given the other. To determine the underdetermination of the system under this combined set of constraints thus involves characterizing the solution set for such a bilinear equation system. Unfortunately, no efficient solution methods for large bilinear equation systems are known, except in certain special cases (Cohen and Tomasi, 1997; Johnson and Link, 2009).

Nevertheless, one can attempt to solve the system by

minimizing the objective function

$$C(\mathbf{B}, \mathbf{T}) = \|\mathbf{K}_1 \text{vec}(\mathbf{B}) - \mathbf{k}_1\|^2 + \qquad (4)$$
$$\|\mathbf{K}_2 \text{vec}(\mathbf{T}) - \mathbf{k}_2\|^2 + \|\mathbf{T}(\mathbf{I} - \mathbf{B}) - \mathbf{I}\|^2,$$

where the $\|\cdot\|^2$ denotes squared Euclidean norm (for a vector) and squared Frobenius norm (for a matrix), i.e. in both cases the sum of squares of the elements. In this objective function, the first term (involving $\mathbf{K}_1$ and $\mathbf{k}_1$) represents all linear constraints on the elements of $\mathbf{B}$ that we have been able to construct, the second term (involving $\mathbf{K}_2$ and $\mathbf{k}_2$) similarly represents any available linear constraints on the elements of $\mathbf{T}$, while the last term derives from the constraint $\mathbf{T} = (\mathbf{I} - \mathbf{B})^{-1}$. This objective function is quadratic in $\mathbf{B}$ (for fixed $\mathbf{T}$), and quadratic in $\mathbf{T}$ (for fixed $\mathbf{B}$), so it is easy to set up an alternating variables approach which minimizes with respect to each one at a time, keeping the other fixed. Since the objective is not convex with respect to the joint parameter vector, one cannot be guaranteed to find the global minimum using such a procedure. In our experience, however, this approach is generally quite effective in finding solutions to the underlying system of equations, when such solutions exist. We use this procedure, starting from a sample of random initial points, to generate a sample of solutions to the available constraints. For each of the direct effects (elements of $\mathbf{B}$) we then, on the basis of its variance in the sample, classify it as determined or underdetermined. For determined coefficients we further infer whether the coefficient is zero (edge is absent) or non-zero (edge is present), based on the mean value of the coefficient in the sample. Note that any constraints on the model parameters that follow the above bilinear form can easily be added to this inference procedure. In fact, it turns out that several of the faithfulness constraints derived in the following section can be added. With these additions, we empirically evaluate this method (under the abbreviation 'BILIN') in the simulations in Section 7.

## 5 FAITHFULNESS CONSTRAINTS ON SETS OF PATHS

While constraints of the form $b(x_i \to x_j) = 0$ are the most obvious consequences of the faithfulness assumption, they by no means exhaust the inferences that are possible in linear models. In this section, we provide a more general framework for deriving and representing faithfulness constraints, and in the following section we make use of these constraints in an inference algorithm for the overlapping data sets setting.

We build on the work of Claassen and Heskes (2011), who developed a logical inference procedure based on *minimal* conditioning sets to derive graphical consequences of the faithfulness assumption. A minimal conditioning set, indicated by the brackets, identifies exactly the set of variables that make a pair of variables (in)dependent. Formally, for variables $x$ and $y$ and disjoint sets of variables $C$ and $D$ not containing $x$ and $y$, we denote a minimal independence by

$x \perp\!\!\!\perp y \,|\, D \cup [C]$ whenever we have
$\quad x \perp\!\!\!\perp y \,|\, D \cup C$ and $\forall C' \subsetneq C, x \not\perp\!\!\!\perp y \,|\, D \cup C'$,

and a minimal dependence by

$x \not\perp\!\!\!\perp y \,|\, D \cup [C]$ whenever we have
$\quad x \not\perp\!\!\!\perp y \,|\, D \cup C$ and $\forall C' \subsetneq C, x \perp\!\!\!\perp y \,|\, D \cup C'$.

In both cases $D$ and $C$ can be empty, although when $C$ is empty, the statements become trivial. Under the assumption that the generating model is acyclic, Claassen and Heskes (2011) provide causal inference rules based on such minimal dependencies and independencies, that identify all consequences of faithfulness given passive observational data on a set of (potentially) causally insufficient variables. We drop their acyclicity axiom and change and expand their rules to apply to overlapping data sets, experimental data, and cyclic generating models. We use them to derive constraints on the experimental effects of hypothetical experiments, which, in turn, are used to identify additional features of the underlying causal model.

For example, if we find that $x \perp\!\!\!\perp y \,|\, [z]$ in the manipulated distribution where only $z$ is subject to an intervention, then we infer that $t(x \rightsquigarrow y || \{x, z\}) = 0$, i.e. for an experiment in which $x$ and $z$ are subject to intervention, there would be no correlation between $x$ and $y$, since, by faithfulness, there is no directed path from $x$ to $y$ that does not pass through $z$. Moreover, we can generalize this constraint to all supersets of the intervention set:

$$t(x \rightsquigarrow y || \mathcal{J}) = 0 \quad \Rightarrow \quad t(x \rightsquigarrow y || \mathcal{J} \cup C) = 0 \qquad (5)$$

for all $C \subseteq \mathcal{V} \setminus \{y\}$. Note, in particular, that $C$ may contain variables in $\mathcal{L}$. As a special case of this rule we thus obtain the direct effect constraint $t(x \rightsquigarrow y || \mathcal{V} \setminus \{y\}) = b(x \to y) = 0$. Similarly, $x$ and $y$ can be reversed in all the previous constraints. If we additionally found for some non-intervened $w$, that $x \not\perp\!\!\!\perp y \,|\, \{z\} \cup [w]$, then $w$ is a so-called "unshielded collider", which implies that $t(w \rightsquigarrow x || \{w, z\}) = t(w \rightsquigarrow y || \{w, z\}) = 0$. Again we can also generate constraints for all the experimental effects with supersets of the intervention sets.

Since faithfulness in general implies the absence of particular causal pathways given a set of discovered independence constraints, it implies for the linear models that we have been considering, that the sum-products of edge coefficients representing a (set of) path(s) connecting two variables are zero. An independence between $x$ and $y$ in a passive observational data set can thus give rise to a set of polynomial equations representing the fact that there is no directed pathway from $x$ to $y$ or vice versa, and that there is no (latent) common cause of $x$ and $y$. Quite apart from the large number of equations such a general approach produces, there are no known efficient solution methods for such a set of equations. We thus have to determine which subset of the consequences of faithfulness we can handle.

We found that for consequences of faithfulness that can be represented as bilinear equations on the experimental effects, such as Equation 6, computationally feasible solution methods could still be developed.

$$t(x{\rightsquigarrow}u||\mathcal{J}\cup\{x\}) \cdot t(u{\rightsquigarrow}y||\mathcal{J}\cup\{u\}) = 0 \quad (6)$$

The bilinear approach of Section 4 can include such equations as (6) if one of the experimental effects is a total effect and the other is a direct effect. In Section 6 we present a method that handles versions of Equation 6 in its more general form. In Appendix D we describe the full set of faithfulness rules we apply and the equations we use. These include the skeleton rules of the PC-algorithm, as well as many other standard faithfulness inferences. We also note the rules' limitations by giving an example of an inference that we cannot include in the current bilinear representation.

The independence constraints due to faithfulness thus allow us to obtain zero-constraints on a variety of experimental effects and their products without ever having to perform the corresponding experiment or having a data set that includes all the variables present in the constraint. We note that in practice inferences due to faithfulness are very sensitive to the correctness of the independence test judgments. In particular, once the inference depends on more than a simple unconditional independence test, reliability rapidly decreases and the theoretical gains one obtains in the infinite sample limit evaporate (see Section 7 for more discussion and comparisons). So while a complete method for inferences based on faithfulness given the background assumptions here is still of theoretical interest, its practical usefulness may be limited.

## 6 INFERENCE ALGORITHM

In Section 4 we took an approach that overparameterized the search problem by treating both the total and the direct effects as unknown. The proposed method searched for solutions using the available constraints that could be represented in the bilinear system relating total and direct effects. But in the previous section we showed that the assumption of faithfulness implies zero-constraints on the experimental effects that cannot, in general, be used by such a bilinear method (see Equations 5 & 6). In this section we propose a method that includes these more general constraints on experimental effects that are neither total nor direct effects. We consider *all* experimental effects as unknown parameters of the model. This can be seen as taking the overparametrization one step further, but given that there are now a total of $n(n-1)2^{n-2}$ different experimental effects for a model with $n$ variables in $\mathcal{V}$, the iterative approach of Section 4 becomes unfeasible; something simpler is needed.

---

**Algorithm 1** Linear Inference algorithm: LININF

Record all observed experimental effects.

Record any experimental effects directly identified as zero by the faithfulness rules.

While TRUE,

    Initialize matrix $\mathbf{A}$ and vector $\mathbf{b}$ as empty. Let $\mathbf{x}$ be the vector of the remaining unknown experimental effects.

    For all experimental effects $t(x_i{\rightsquigarrow}x_u||\mathcal{J}_k)$,

        For all sets $\mathcal{J}_l$ such that $\mathcal{J}_k \subsetneq \mathcal{J}_l \subsetneq \mathcal{V}\setminus\{x_u\}$,

            If Equation 7 is *linear* with respect to the unknown experimental effects, add it to the system $\mathbf{Ax} = \mathbf{b}$.

    For all equations induced by the faithfulness rules,

        If the equation is *linear* with respect to the unknown experimental effects, add it to the system $\mathbf{Ax} = \mathbf{b}$.

    Solve $\mathbf{x}$ from $\mathbf{Ax} = \mathbf{b}$ and determine the uniquely identified elements of $\mathbf{x}$.

    Record all experimental effects corresponding to uniquely identified elements of $\mathbf{x}$, and derive implied experimental effects based on Equation 5.

    Exit if no new experimental effects were recorded.

Output the recorded experimental effects.

---

Some experimental effects are readily observed in the overlapping experiments. We use these observed experimental effects to determine some of the remaining unknown ones. Hyttinen et al. (2012) generalized Equation 2 to show how some experimental effects can be used to constrain others. If two experiments with intervention sets $\mathcal{J}_k$ and $\mathcal{J}_l$ satisfy $\mathcal{J}_k \subsetneq \mathcal{J}_l \subsetneq \mathcal{V}\setminus\{x_u\}$, then the experimental effects of the two experiments are related by the following equation:

$$t(x_i{\rightsquigarrow}x_u||\mathcal{J}_k) = t(x_i{\rightsquigarrow}x_u||\mathcal{J}_l) + \sum_{x_j\in\mathcal{J}_l\setminus\mathcal{J}_k} t(x_i{\rightsquigarrow}x_j||\mathcal{J}_k)t(x_j{\rightsquigarrow}x_u||\mathcal{J}_l) \quad (7)$$

Generally such equations are non-linear in the unknown experimental effects. However, the equation is *linear* if for all $x_j \in \mathcal{J}_l \setminus \mathcal{J}_k$ either $t(x_i{\rightsquigarrow}x_j||\mathcal{J}_k)$ or $t(x_j \rightsquigarrow x_u || \mathcal{J}_l)$ is already known. In addition, the full set of implications from the faithfulness rules of Section 5 (see Appendix D) may require that a particular experimental effect is zero, or – as in Equation 6 – that $t(x_i \rightsquigarrow x_j || \mathcal{J}_k) t(x_j \rightsquigarrow x_u || \mathcal{J}_l)$ is zero. Combining all the available information can render several equations linear. We ignore any equations that remain non-linear in the unknown experimental effects in order to keep the solvability of the system feasible.

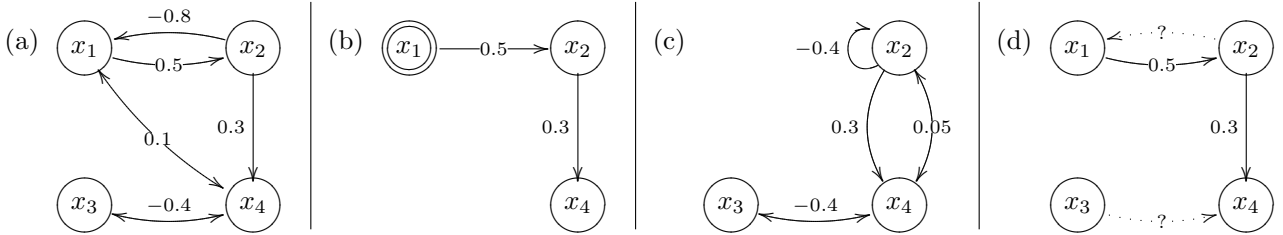

Figure 2: A case study. (a) True data generating model. (b) The manipulated and marginalized model corresponding to an experiment with $\mathcal{J}_1 = \{x_1\}$, $\mathcal{U}_1 = \{x_2, x_4\}$, and $\mathcal{L}_1 = \{x_3\}$. (c) The marginalized model corresponding to a 'null'-experiment with $\mathcal{J}_2 = \emptyset$, $\mathcal{U}_2 = \{x_2, x_3, x_4\}$, and $\mathcal{L}_2 = \{x_1\}$. (d) The causal structure over the union of observed variables learned by LININF (omitting any double-headed arcs). Notice that all causal arcs between $x_1$ and $x_3$ are discovered to be absent, although variables $x_1$ and $x_3$ are not observed together in either of the observed data sets. Dotted arrows indicate underdetermined parameters.

There are a total of $n(n-1) \sum_{i=0}^{n-2} \binom{n-2}{i}(2^{n-i-2} - 1)$ equations similar to Equation 7 for a model with $n$ variables in $\mathcal{V}$. Thus, if even a small share of these equations are linear with respect to the unknown experimental effects, the model space compatible with the constraints at hand will be substantially restricted. Furthermore, equations that were initially non-linear in the unknown experimental effects, can be rendered linear as we iterate the inference and more experimental effects are determined. Since we are only using linear equations, we can trivially see which of the unknown experimental effects are uniquely determined by the system of equations. Similarly, since the direct effects are just a special case of the experimental effects, we can find which direct effects are underdetermined.

The Linear Inference algorithm using these ideas is outlined in Algorithm 1. A powerful part of the algorithm is the application of the inference in Equation 5 throughout the several rounds of inference. For finite samples this inference depends on judging whether an inferred experimental effect that is not measured in any of the available data sets, is zero. One option is to run the same algorithm in parallel on several resampled data sets, and see whether all estimates for a single experimental effect are effectively zero or not.

Unlike the bilinear procedure of Section 4, this linear inference procedure is consistent. It is incomplete as it does not exploit the available non-linear equations, but it is able to extract more information from the overlapping experiments by benefiting from the vast number of different faithfulness constraints produced by sparse data generating models.

## 7 SIMULATIONS

In the overlapping data set setting the general idea is that the experiments have already been run, with little regard to how the data sets and interventions may fit together. So it is likely that the identifiability condition of Section 3 is not satisfied and that there still is a large amount of underdetermination. In particular, one might expect substantial underdetermination with regard to the causal relations between variables that are never measured together in one data set. Figure 2 offers a concrete example of how the findings from multiple experimental data sets can be synthesized to yield knowledge not available from any of the individual data sets. Panel (a) shows the true underlying data generating model. This model is subject to one experiment where $x_1$ is intervened, $x_2$ and $x_4$ are observed, while $x_3$ is latent (b), and one 'null'-experiment where all variables except $x_1$ are passively observed (c). Note that $x_1$ and $x_3$ are never measured together in the same data set. One can think of the two experiments as resulting from two different research groups investigating two overlapping sets of variables. Panel (b) represents the manipulated and marginalized version of the true model corresponding to the first experiment, while (c) represents the marginalized model corresponding to the 'null'-experiment. Panel (d) shows the output of the Linear Inference algorithm described in Section 6 applied to the overlapping data sets of the two experiments. Here, solid edges denote identified edges, absences of edges denotes cases where the algorithm has inferred that there is no edge, and what remains undetermined is shown by dotted edges. Note that faithfulness has been effectively used to detect the absence of several edges in the underlying model. In particular, $x_4$ is determined *not* to be a direct cause of any other variable, without there being any experiment intervening on $x_4$. Also, similarly to the results of Tillman et al. (2009) for passive observational data from acyclic models, we can detect the absence of edges between $x_1$ and $x_3$, *even though they were never measured together in any of the data sets.* Only the existence of the arcs $x_2 \to x_1$ and $x_3 \to x_4$ (indicated by dotted arrows) is left underdetermined. Finally, note that we have not discussed the detection of confounding here, so the absence of bi-directed edges in panel (d) is not indicative of any prediction on such edges.

We assessed the performance of our methods more generally in simulations. To test the effect of the faithfulness assumption we ran the method of Eberhardt et al. (2010) (EHS), that was shown in Section 3 to be complete for our circumstances, as a baseline for comparison. It does not assume faithfulness. We compared this baseline with the performance of the new BILINear and LINear-INFerence methods, presented in Sections 4 and 6, respectively, that were specifically designed for the overlapping data set case, and that assume faithfulness. Lastly, we ran the method of Hyttinen et al. (2010) (HEH), that only uses faithfulness constraints on the *direct effects* obtained *within* each data set.

The methods[3] were modified to produce one of three answers: 'present', ' absent', or 'unknown', for each possible edge, i.e. each entry of **B**. An algorithm could only return 'present' if it had actually identified a specific value for the edge parameter. We compared how much of the true model the algorithms identified, and what percentage of that was correct.

To ensure that assuming faithfulness could actually make a difference, we simulated the methods on sparse models. We generated 100 6-variable models, with around 20% of the possible edges present, and 15% of the possible non-zero covariances, representing exogenous latent confounders. We drew the connection strengths randomly, though we avoided generating coefficients that were very close to zero, in order to make the predictions of absence/presence of edges meaningful. For each model, we constructed 5 overlapping experiments by selecting uniformly among partitions of the set of variables into intervened, observed and hidden variables, with the only constraint that for each experiment, neither $\mathcal{J}$ nor $\mathcal{U}$ were empty.

Figure 3 reports the percentage correct among the edges reported absent (left) and present (right) for each algorithm at 1,000 and 10,000 samples per experiment, as well as for the infinite sample limit. Note that for the finite sample cases we restricted the maximum size of the conditioning set, in the search for faithfulness rules, to 1; this tends to improve the accuracy of the predictions. Similarly, in the finite sample cases, we restricted the number of inference rounds in the LININF algorithm to 1. With the exception of BILIN, the algorithms have a similar performance across the board. BILIN offers an undesirable trade-off: for finite samples it offers a high correctness of predicted absences, but a poor performance for predicted links. And since it is not consistent, it does not guarantee the absences of errors in the infinite sample limit, as can be seen in the presence of errors for the predicted absences.

---

[3]Code implementing all compared methods and reproducing the simulations is available at
http://www.cs.helsinki.fi/u/ajhyttin/exp/

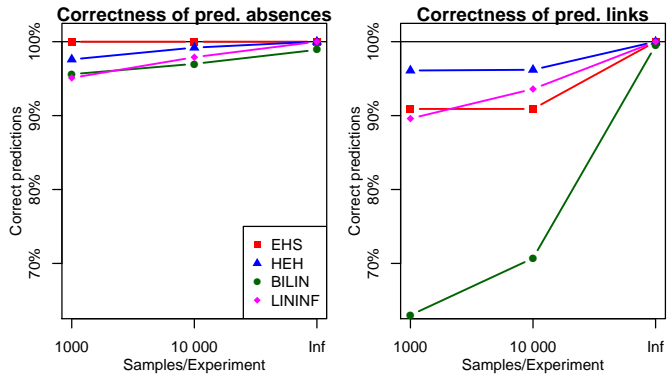

Figure 3: Accuracy of the learning algorithms. Each point on the lines is the average over 100 models.

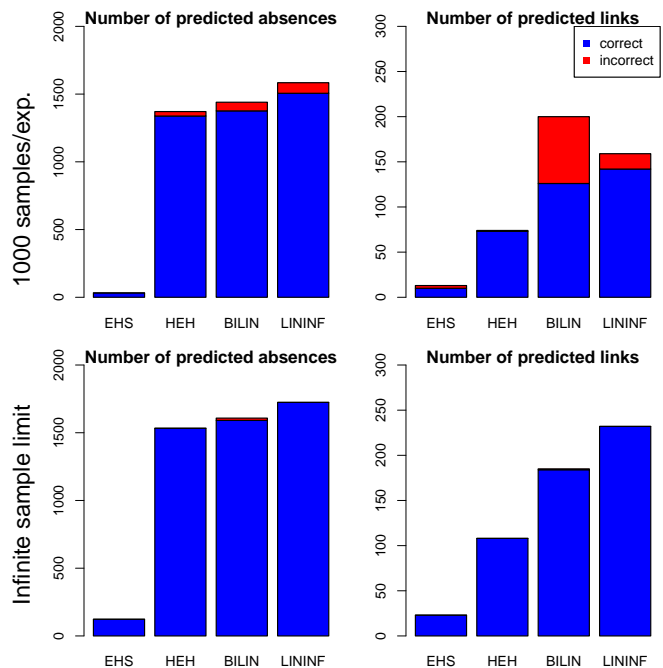

Figure 4: Predictions of the learning algorithms. Each bar represents the predictions over 100 models.

Figure 4 shows the actual numbers of correct and incorrect predicted absences (left column) and links (right column) for each algorithm at 1,000 samples per experiment (top row) and the infinite sample limit (bottom row). In the 100 true models of the simulation there were a *total* of 624 edges and 2376 absences (confounding is not included in either of these counts). The first thing to notice is the scale of the plots: Substantial underdetermination remains for all algorithms, especially with respect to the identified links: More than half the links remain underdetermined. With respect to the individual algorithms we notice that EHS produces very few predictions, most edges are simply marked as 'unknown', since faithfulness is not assumed. HEH produces a remarkable number of predicted absences with good accuracy, so that the improvement of the methods tailored to the setting of overlapping experiments is limited for finite samples. However, the improvement in the number of predicted links is clearly quite significant (top and bottom right plots).

It would be desirable to have a comparison of how much underdetermination is inevitable given the overlapping data sets. A brute force check is not feasible even for six variable models, and we are not aware of any general results on equivalence classes for (manipulated) linear cyclic models with latent confounders. Developing an extension of LININF that is complete with regard to faithfulness may thus be the more promising route for such a comparison.

## 8 CONCLUSION

We have presented two new algorithms (BILIN and LININF) designed for causal learning from overlapping experimental or passive observational data sets. The first algorithm relies on a bilinear iterative optimization, while the second is based on solving linear constraints on general experimental effects. Both algorithms assume faithfulness, but apply to the general model class of linear cyclic models with latent confounding. We have also formulated necessary and sufficient conditions for model identifiability when faithfulness is not assumed.

The approach we have taken brings together several different strands of related work: We have connected the results on combining different *experimental* results on the *same* set of variables with the techniques of integrating *overlapping* data sets of *passive observational* data. To do so, we have relied on the inferences based on faithfulness developed for *non-parametric, acyclic* causal models with latent confounding, but adapted them to *linear, cyclic* models with latent confounding.

### Acknowledgements

The authors would like to thank the anonymous reviewers for their comments and valuable suggestions. F.E. was supported by a grant from the James S. McDonnell Foundation on 'Experimental Planning and the Unification of Causal Knowledge'. A.H. and P.O.H. were supported by the Academy of Finland.

### References

Bollen, K. A. (1989). *Structural Equations with Latent Variables*. John Wiley & Sons.

Claassen, T. and Heskes, T. (2011). A logical characterization of constraint-based causal discovery. In *UAI 2011*.

Cohen, S. and Tomasi, C. (1997). Systems of bilinear equations. Technical report, Stanford University.

Cooper, G. and Yoo, C. (1999). Causal discovery from a mixture of experimental and observational data. In *UAI 1999*.

Eaton, D. and Murphy, K. (2007). Exact Bayesian structure learning from uncertain interventions. In *AISTATS 2007*.

Eberhardt, F., Hoyer, P. O., and Scheines, R. (2010). Combining experiments to discover linear cyclic models with latent variables. In *AISTATS 2010*.

He, Y. and Geng, Z. (2008). Active learning of causal networks with intervention experiments and optimal designs. *Journal of Machine Learning Research*, 9:2523–2547.

Hyttinen, A., Eberhardt, F., and Hoyer, P. O. (2010). Causal discovery for linear cyclic models. In *PGM 2010*.

Hyttinen, A., Eberhardt, F., and Hoyer, P. O. (2011). Noisy-or models with latent confounding. In *UAI 2011*.

Hyttinen, A., Eberhardt, F., and Hoyer, P. O. (2012). Learning linear cyclic causal models with latent variables. Submitted. Available online from the authors' homepages.

Johnson, C. R. and Link, J. A. (2009). Solution theory for complete bilinear systems of equations. *Numerical Linear Algebra with Applications*, 16(11-12):929–934.

Murphy, K. P. (2001). Active learning of causal Bayes net structure. Technical report, U.C. Berkeley.

Pearl, J. (2000). *Causality: Models, Reasoning, and Inference*. Cambridge University Press.

Schmidt, M. and Murphy, K. (2009). Modeling discrete interventional data using directed cyclic graphical models. In *UAI 2009*.

Spencer, J. (1970). Minimal completely separating systems. *Journal of Combinatorial Theory*, 8(4):446 – 447.

Spirtes, P., Glymour, C., and Scheines, R. (1993). *Causation, Prediction, and Search*. Springer-Verlag.

Tillman, R. E., Danks, D., and Glymour, C. (2009). Integrating locally learned causal structures with overlapping variables. In *NIPS 2008*.

Tillman, R. E. and Spirtes, P. (2011). Learning equivalence classes of acyclic models with latent and selection variables from multiple datasets with overlapping variables. In *AISTATS 2011*.

Tong, S. and Koller, D. (2001). Active learning for structure in Bayesian networks. In *UAI 2001*.

Triantafillou, S., Tsamardinos, I., and Tollis, I. G. (2010). Learning causal structure from overlapping variable sets. In *AISTATS 2010*.